\documentclass[amssymb,nobibnotes,superscriptaddress,longbibliography,twocolumn,floatfix]{revtex4-1}

\usepackage{enumerate,amsmath}
\usepackage{graphicx}
\usepackage{color}

\begin{document}
\title{Physical reservoir computing---An introductory perspective}%
\author{Kohei Nakajima}
\email{k\_nakajima@mech.t.u-tokyo.ac.jp}
\address{Graduate School of Information Science and Technology, The University of Tokyo, Bunkyo-ku, 113-8656 Tokyo, Japan}
\begin{abstract}
Understanding the fundamental relationships between physics and its information-processing capability has been an active research topic for many years. 
Physical reservoir computing is a recently introduced framework that allows one to exploit the complex dynamics of physical systems as information-processing devices. 
This framework is particularly suited for edge computing devices, in which information processing is incorporated at the edge (e.g., into sensors) in a decentralized manner to reduce the adaptation delay caused by data transmission overhead. 
This paper aims to illustrate the potentials of the framework using examples from soft robotics and to provide a concise overview focusing on the basic motivations for introducing it, which stem from a number of fields, including machine learning, nonlinear dynamical systems, biological science, materials science, and physics.
\end{abstract}

\maketitle
\section{Introduction}
Recently, a novel information-processing scheme that exploits physical dynamics as a computational resource has been proposed.
This scheme is called {\it physical reservoir computing} (PRC).
The current paper aims to introduce this framework concisely, focusing on its motivation and potential by using a number of examples.
Understanding the original concept of reservoir computing (RC) is important to comprehend the concept of PRC.
RC is a framework for recurrent neural network (RNN) training and was proposed in the early 2000s as a broad concept that allows to deal with a number of different models of RNN, including the echo-state network (ESN) \cite{Jaeger1,Jaeger_tutorial,Jaeger2} and the liquid state machine (LSM) \cite{Maass1}, under the same umbrella \cite{Schrauwen1,Schrauwen2,Jaeger3,Jaeger4}.

Conventionally, to train an RNN, a backpropagation-through-time (BPTT) method \cite{BPTT} is frequently used.
In this method, all the weights of the network are basically tuned toward the target function.
In the RC framework, by preparing an RNN equipped with a massive amount of nonlinear elements coupled with one another, called a {\it reservoir}, only the readout part is usually trained toward the target function.
In the simplest case, this readout part consists of linear and static weights that directly connect the reservoir nodes and output node (Fig. \ref{fig1}A).
Because of this unique system construction, RC has many advantages.
Some typical examples are given below.

\begin{figure*}
\includegraphics[width=180mm]{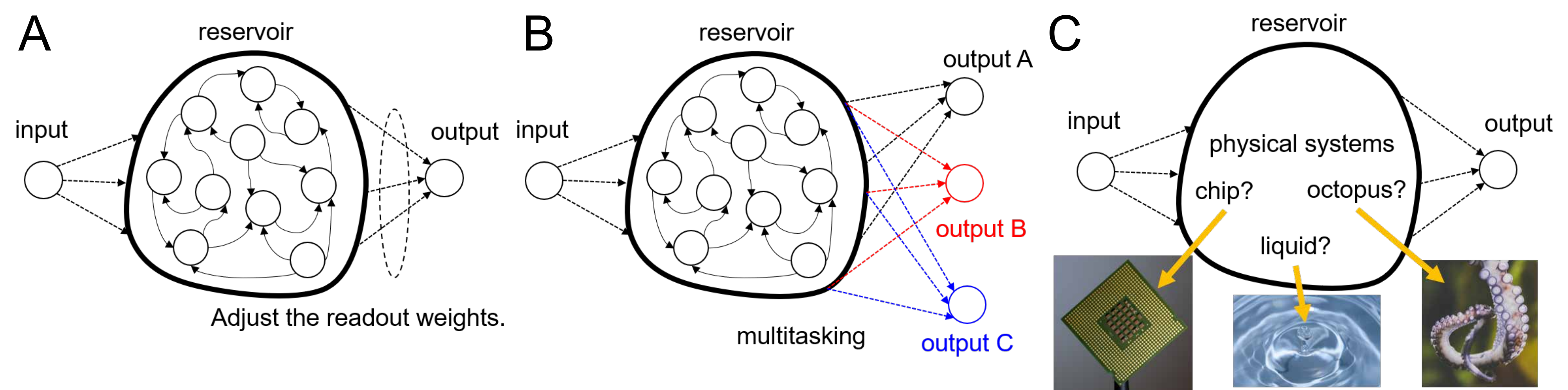}
\caption{{\bf Typical settings and advantages in RC.} 
A. A typical ESN setting, a representative model in the RC framework.  
The reservoir is an RNN often equipped with a nonlinear activation function, such as $y=\tanh(x)$.
Only the readout part is usually trained to the target function.
B. In the RC framework, multitasking can be safely implemented in principle, because no interference occurs among the tasks during the learning procedures. 
See the text for details.
C. Physical reservoir computing, which exploits the physical dynamics as a reservoir.
}
\label{fig1}
\end{figure*}

The first advantage comes from the ease in the training procedure, which makes the learning quick and stable. 
As noted above, in the conventional BPTT approach, all the weights in the network are tuned, which takes a significant amount of time in obtaining the optimal parameter set according to the type of the given target function. 
Furthermore, it is known to be unstable, in general, suggesting that it cannot always obtain the optimal set of weights after learning \cite{BPTT_difficult}. 
In the RC framework, the weights in the network are not always targeted for training.
Instead, the training is mainly for the readout part, so the number of parameters that need to be tuned is generally small, making the training significantly faster (Fig. \ref{fig1}A). 
In particular, if the readout part is set as linear and static weights, the training can be executed with a simple linear regression or ridge regression, and the optimal set of weights can be induced at once through a batch learning procedure, making the entire learning process simple and stable. 
Accordingly, there are many real-world application scenarios proposed in the literature. 
Starting from conventional signal processing for robust communication against noise \cite{Jaeger2}, learning of the grammatical structure of natural language \cite{ESN_Grammer}, robust speech recognitions \cite{ESN_SpeechRecognition}, or handwritten digit recognitions \cite{ESN_HandwrittenRecognition}, many attempts can be found for complex time series prediction tasks, including the time series of stock markets \cite{ESN_Stockmarket1,ESN_Stockmarket2} or for the prediction of high-dimensional spatiotemporal dynamics found in nature \cite{Ott_PRL}, including weather forecasting or the prediction of forest fire spreading. 
In robotics, for example, many cognitive tasks, which were previously difficult to implement because of the complicated procedure of RNN training, have been revived using RC for cognitive agents \cite{ESN_Cognitive}, and behavioral generations of robots, such as the emulation of motor controller \cite{ESN_Motorcontrol,Li_ICMA2012,Li_ICRA2012,Li_ICRA2013},  inverse kinematics \cite{ESN_Inverse}, timing control \cite{Kuwabara_IJCNN2012}, and central pattern generator (CPG) \cite{ESN_CPG}, are successfully performed. 
In addition, researchers are now interested in applying the RC framework to sensory devices, in which the raw data are collected, and for executing processing natively on the sensory devices in real time, which is called edge computing \cite{Edge_Computing}. 
Fonollosa et al. applied an RC framework to a chemical gas sensory system and showed that it is suitable for real-time and continuous monitoring applications and improves the time response of the chemical sensory system \cite{ESN_Gassensor}. 
Recently, the emulation of the functionality of a sensory device in a soft robotic platform was proposed using ESN, where the laser displacement sensor is emulated in a significantly high accuracy \cite{Robosoft2020}. 
This approach is expected to replace the functionality of rigid components, that is, sensory devices, freeing soft robotic platforms from mechanical constraints to maintain their softness and flexibility. 
We should note that although the learning procedure of RC is simple, this does not imply that RC is less powerful than conventional machine learning techniques \cite{Convex_learning}. 
For example, it has been shown that ESN, which is a representative model system of RC, has a universal approximation property, and many studies are now proving its expressive power in different settings \cite{RC_Universality1,RC_Universality2}. 
This implies that it is largely up to the experimenters using the framework and how they will utilize it to induce its potential. 
In a machine learning context, many improvements have been proposed to overcome the instability of RNN learning based on BPTT algorithms, which can be represented in the model of long-short term memory \cite{LSTM}, gated recurrent unit \cite{GRU}, or unitary RNNs \cite{Unitary_RNN1,Unitary_RNN2}. 
Among these approaches, a recent systematic comparison analysis with RC has shown that each of these approaches has its merits and demerits (see Ref.\cite{RC_BPTT_Comparison} for more details), which suggests that the best approach depends on the experimental conditions and is largely up to what the experimenters wish to achieve. 

The second advantage is its ease in multitasking or in sequential learning. 
Consider that the network is now implementing a task $T_A$ to the output $A$ according to the input $u$, which is expressed as $y_A=T_A (u)$. 
Now, we want to train the same network to additionally learn the task $T_B$ to the output $B$ according to the same input $u$, which is expressed as $y_B=T_B (u)$. 
In the conventional approach of backpropagation, the entire network is optimized for the task $T_A$ first, and then the network is additionally trained for the task $T_B$ using the backpropagation method, so these two tasks interfere during the update of weights within the same network. 
In this situation, there is danger that the network forgets the previously learned tasks. 
The extreme case for this phenomenon is called {\it catastrophic interference} or {\it catastrophic forgetting} \cite{Catastrophic_forgetting1,Catastrophic_forgetting2}, and addressing this deficit remains a controversial topic for many researchers (see, e.g., Ref. \cite{Catastrophic_forgetting_solve1,Catastrophic_forgetting_solve2,Catastrophic_forgetting_solve3}). 
In the RC framework, because the training is basically limited at the readout part, no interference occurs among the tasks, so multitasking can safely be implemented in principle (Fig. \ref{fig1}B).

The third advantage is the arbitrariness and diversity in the choice of a reservoir. 
The basic concept of RC is exploiting the intrinsic dynamics of the reservoir by outsourcing learning, which requires some parameter tuning, to the readout part. 
According to this unique setting, reservoirs do not have to be an RNN anymore but can be any dynamical system. 
This idea naturally leads us to exploit the physical dynamics as a reservoir instead of using the simulated dynamics inside the PC (Fig. \ref{fig1}C). 
This framework is called PRC and is a main theme of the current paper. 
This seemingly natural step makes the framework radically different from other machine learning methods. 
That is, PRC provides a novel insight not only into the machine learning community, but also into the dynamical systems field, physics, materials science, and biological science. 
This point will be elaborated on in detail later.

\section{Prerequisite for a successful reservoir}
As we verified in the previous section, there is a diversity in the choice of reservoir, and there is a freedom to use any kind of dynamical system if you wish. 
However, whether that reservoir works successfully is a different story. 
There exists a prerequisite to be used as a successful reservoir. 
The prerequisite is about the reproducibility of the input--output relation, which is an inevitable condition for any computational device. 
Namely, the reservoir should respond the same whenever the same input sequence is injected. 
Otherwise, every time you used it, the reservoir would respond differently, meaning that it would be operationally troublesome and unreliable. 
Considering that the reservoir is basically a dynamical system, this requirement is a somewhat severe condition because the behavior of dynamical systems is in general determined by the initial condition. 
If you can precisely select the initial condition of the reservoir and can control the timing to inject the input sequence into the system, then for the identical input sequence, you can always obtain the same response from the system.
However, this constraint restricts the usability of the computational device, and it is particularly annoying if you wish to exploit the natural and physical dynamics as a reservoir because it is generally difficult to infer or control the initial condition of the physical dynamics.
It is preferable to guarantee the reproducibility of the response whenever you inject the same input sequence and, furthermore, to do so without controlling the initial condition of the reservoir.
The property that realizes these conditions of the reservoir is called the {\it echo state property} (ESP) \cite{Jaeger1}.
Simply put, ESP requires the reservoir states to be expressed as a function of the previous input sequence only.
A similar concept has been studied in the nonlinear dynamical systems field from a different angle as a synchronization phenomenon between two identical systems induced by a common signal (or noise) or a generalized synchronization between an input sequence and the corresponding response of the system (see, e.g., Ref. \cite{Toral_NIO}).
This property suggests that even if the system is driven by a different initial condition, by injecting an input sequence, the corresponding response of the system becomes the same.
Mathematical investigations of the concept of ESP (e.g., Ref. \cite{ESP1,ESP2,Yaguchi2}) and understanding its relation to the nonlinear dynamical systems field are still ongoing research topics (e.g., Ref. \cite{RC_dynamics}).

\begin{figure*}
\includegraphics[width=160mm]{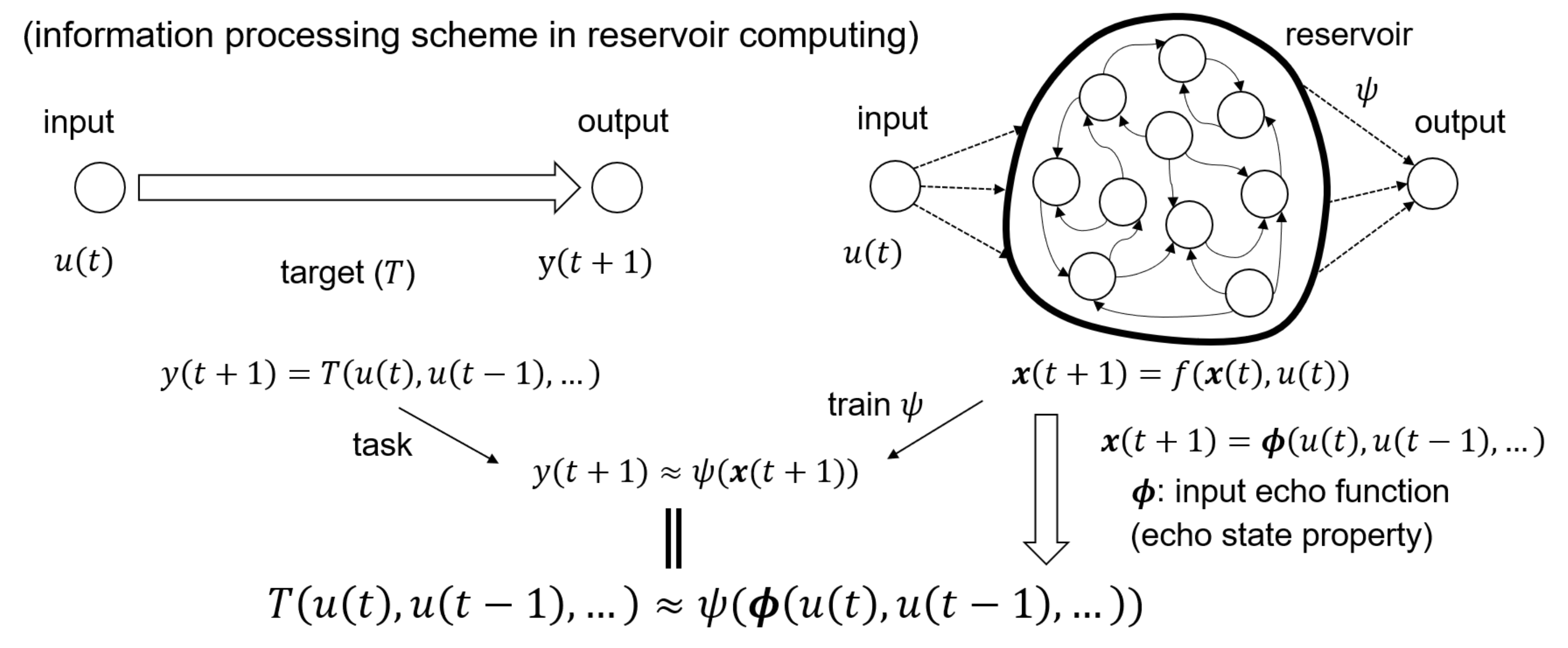}
\caption{{\bf Schematics showing how the echo state property works in RC.} As can be seen in the diagram, input echo function $\mbox{\boldmath $\phi$}$ is a part intrinsic to the reservoir, and experimenters can adjust the output using readout function $\psi$. See the text for details.}
\label{fig2}
\end{figure*}

Here, we would like to summarize the situation briefly (Fig. \ref{fig2}). 
Consider that we have the input $u(t)$ and the reservoir state $\mbox{\boldmath $x$}(t)$ at timestep $t$, and the reservoir dynamics is expressed as $\mbox{\boldmath $x$}(t+1)=f(\mbox{\boldmath $x$}(t),u(t))$. 
In general, a task $T$ targeted by RNN is a function of the previous input sequence, which is sometimes called a temporal machine learning task; then, it is expressed as $y(t+1)=T(u(t),u(t-1),...)$. 
In the RC scheme, by tuning the readout $\psi$ (note that this readout function does not have to be linear in general), we aim to approximate the target $y(t)$, which is expressed as $y(t) \approx \psi (\mbox{\boldmath $x$}(t))$. 
Now, if the reservoir fulfils the ESP, then $\mbox{\boldmath $x$}(t)=\mbox{\boldmath $\phi$}(u(t-1),u(t-2),..)$, where $\mbox{\boldmath $\phi$}$ is called the {\it input echo function} in Ref. \cite{Jaeger1} and where it is a function intrinsic to the reservoir. 
This implies that the internal state of the reservoir is completely described by the driven input sequence and is related to the filter concept, which will be discussed in more detail later. 
Note that when the ESP holds, then the reservoir states from different initial conditions, which are expressed as $\mbox{\boldmath $x'$} (t)$ and $\mbox{\boldmath $x$}(t)$ and driven by identical input sequence, will respond the same or become synchronized, such as $|f(\mbox{\boldmath $x'$}(t),u(t))-f(\mbox{\boldmath $x$}(t),u(t))| \approx 0$ for a sufficiently large $t$. 
In summary, the RC scheme can be expressed as exploiting the function intrinsic to the reservoir $\mbox{\boldmath $\phi$}$ and adjusting the readout function $\psi$ to approximate the target function $T$, which is expressed as $T(u(t),u(t-1), ...) \approx \psi(\mbox{\boldmath $\phi$} (u(t),u(t-1), ...))$.           

From this viewpoint, evaluating the information processing capability or expressive power of a given reservoir is nothing but evaluating the property of the function $\mbox{\boldmath $\phi$}$. 
Currently, several approaches exist. 
The typical case is evaluating how well the given reservoir can output the previous input sequence, and this measure is called memory capacity \cite{Jaeger_MC}. 
Focusing on ESN, the behaviors of memory capacity and their related measures are studied in detail with a linear activation function \cite{Jaeger_MC,White_MC,Ganguli_MC,Hermans_MC,Tino_MC,Goudarzi_MC} and, recently, with a nonlinear activation function \cite{Toyoizumi_MC,Schuecker_MC,Haruna_MC}. 
This measure is further generalized and extended to be able to evaluate the nonlinear memory capacities by decomposing the function $\mbox{\boldmath $\phi$}$ into the combinations of multiple orthogonal polynomials \cite{Dambre}, and the trade-off between the expressiveness of $\mbox{\boldmath $\phi$}$ for linear and nonlinear functions is investigated \cite{Dambre,Inubushi}.
Investigations of the relationships between the dynamical property of the reservoir and its information processing capability are now ongoing hot topics in the field \cite{Stepney_RC}.
Discussions that include how the bifurcation structure or the order-chaos transition (the critical point is often referred to as {\it edge of chaos}) affects the computational power of the reservoir are one such example \cite{Edge_of_chaos1,Edge_of_chaos2,Edge_of_chaos3,Edge_of_chaos4}.

As we confirmed in this section, although, on the one hand, the learning procedure seems simple in RC, which is outsourced to the readout part, on the other hand, the reservoir part can be taken as a huge hyper parameter that is difficult to harness without knowledge of nonlinear dynamical systems. 

\section{Diverse variations of reservoir: toward exploiting physical dynamics}
\begin{figure*}
\includegraphics[width=170mm]{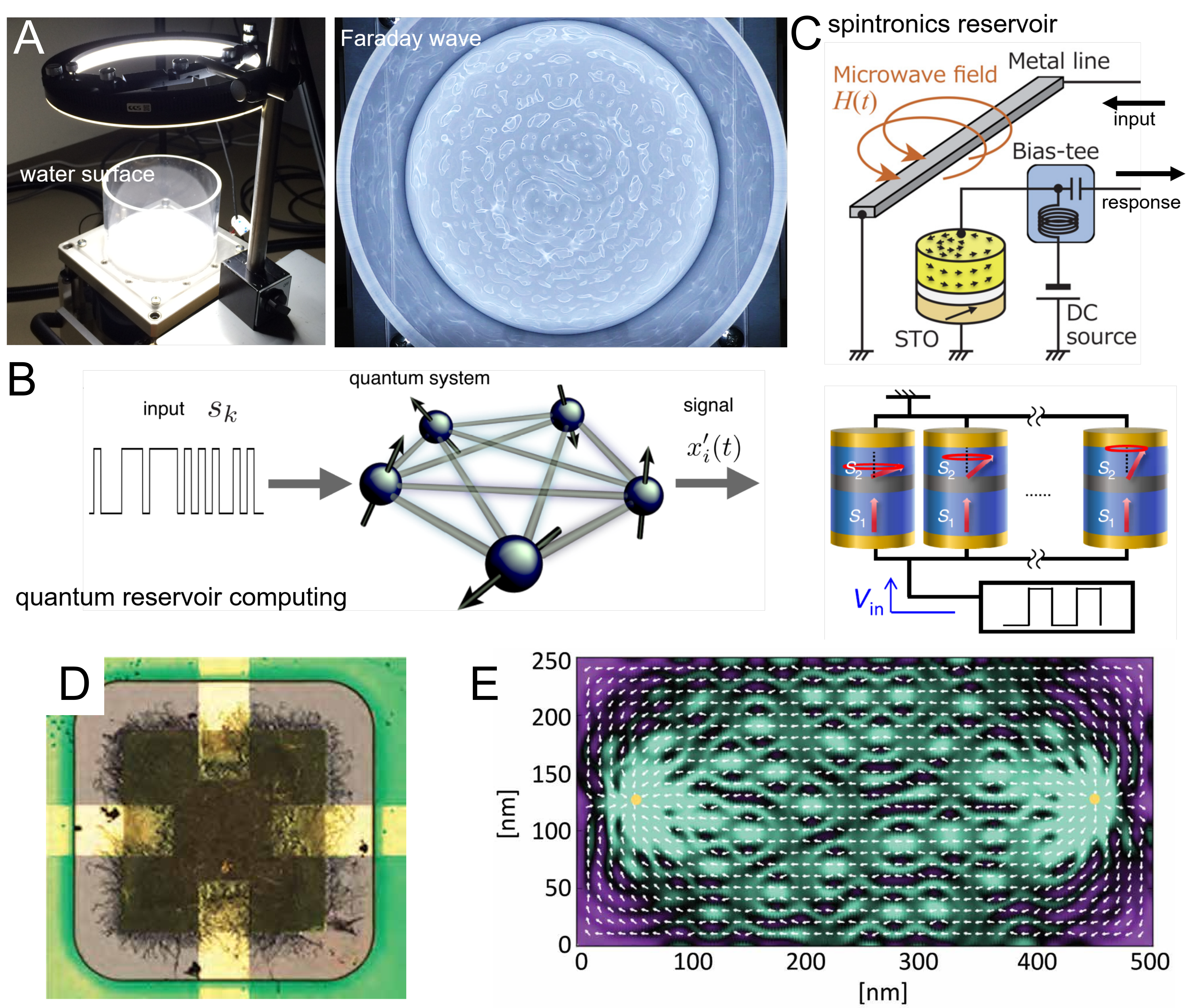}
\caption{{\bf Variations of physical reservoirs.} 
A. The physical liquid state machine proposed in Ref. \cite{Fluid_Nakajima}. 
It exploits the Faraday wave as a computational resource. 
B. Quantum reservoir computing proposed in Ref. \cite{QR_Nakajima1}. 
It allows to exploit disordered ensemble quantum dynamics as a computational resource.
Figure reprinted with permission from Ref. \cite{QR_Nakajima1}, Copyright (2017) by the American Physical Society.
C. Variations of the spintronics reservoir. 
The upper and lower diagrams show reservoirs, which exploit vortex-type spintronics \cite{Spin_Tsunegi2} and spatially multiplexed magnetic tunnel junctions \cite{Spin_Furuta_PRAppl}, respectively.
The upper diagram is a figure reprinted with permission from Ref. \cite{Spin_Tsunegi2} by the author.
The lower diagram is a figure reprinted with permission from Ref. \cite{Spin_Furuta_PRAppl}, Copyright (2018) by the American Physical Society.
D. Complex Turing B-type atomic switch networks proposed in Ref. \cite{Nano_Atom1}.
The complex nanowire network extends throughout the device and is probed via macroscopic electrodes.
Figure reprinted with permission from Ref. \cite{Nano_Atom1}, Copyright (2012) by John Wiley and Sons.
E. A skyrmion network embedded in frustrated magnetic films proposed in Ref. \cite{Spin_skyrmion}.
The current path is visualized after the voltage is applied to the frustrated magnetic texture including Bloch skyrmions.
Figure reprinted with permission from Ref. \cite{Spin_skyrmion}, Copyright (2018) by the American Physical Society.
}
\label{fig3}
\end{figure*}

As we discussed in the previous sections, many types of reservoirs are now proposed.
Among these, a framework that exploits the physical dynamics as the reservoir is called PRC.
Because the natural physical dynamics is directly used as a computational resource, even if the same computation is implemented, according to the different physical property, there will be diverse application scenarios.
Increasingly, many physical reservoirs have been reported worldwide (Fig. \ref{fig3}), such as the case using
photonics \cite{Optic_reservoir_Vandoorne1,Optic_reservoir_Larger,Optic_reservoir_Paquot,Optic_reservoir_Duport,Optic_reservoir_Brunner,Optic_reservoir_Vandoorne2,Optics_Uchida,Optics_Larger,Optics_Brunner,Optics_Bueno,Optics_Takano},
spintronics \cite{Spin_Julie1,Spin_Furuta_PRAppl,Spin_Tsunegi1,Spin_Nakane1,Spin_Tsunegi2,Spin_Julie2,Spin_Nomura1,Spin_Kanao,Spin_Julie3,Spin_Liu,Spin_Nomura2,Spin_Julie4},
quantum dynamics \cite{QR_Nakajima1,QR_Nakajima2,QR_Ghosh1,QR_Ghosh2,QR_Ghosh3,QR_Yamamoto},
nanomaterials \cite{Nano_Atom1,Nano_Atom2,Nano_Susan1,Nano_Susan2,Nano_Memristor1,Nano_Atom3,Nano_Akai2,Nano_Memristor2,Nano_Memristor3},
analog circuits and field programmable gate arrays \cite{Analog_Appeltant1,Analog_Appeltant2,Analog_Alomar1,Analog_Alomar2,Analog_Antonik1,Analog_Alomar3,Analog_Alomar4,Analog_Antonik2},
mechanics \cite{Mech_Helmut1,Mech_Helmut2,Mech_Ken1,Mech_Nakajima1,Mech_Nakajima2,Mech_Qian,Mech_Nakajima3,Mech_Ken2,Mech_Nakajima4,Mech_Nakajima5,Mech_Coulombe,Mech_Wyffels,Mech_Yamanaka,Mech_Nakajima6,Mech_Nakajima7,Mech_Isuru},
fluids \cite{Fluid_Maass,Fluid_Fernando,Fluid_Nakajima,Fluid_Goto},
and biological materials \cite{Bio_Maass,Bio_Dranias,Bio_Kubota}.
Readers interested in which types of reservoirs are currently proposed can refer to, e.g., Ref. \cite{Tanaka_PRC,Nakajima_RCBook}.

Before going into PRC, which is the main theme of the current paper, we would like to overview the typical misapprehensions that we frequently face when it comes to the RC framework in this section.
The first one is the belief that the weights of the reservoir should be set randomly. 
Of course, there exists a reservoir that implements a random weight matrix, such as ESN, but this is not an essential requisite.
RC was originally inspired by the type of information processing that occures in the brain, and the connections between neurons are usually not random but have specific structures.
Accordingly, several reservoir settings implement brain-inspired connections \cite{Kawai_NN} or simply implement the neighboring connections \cite{Maass1}, introducing a spatial dimension that is not random at all.
More coherent network structures, such as cyclic reservoirs, are also investigated \cite{Tino_MC}.
One interesting aspect of RC is that it is capable of exploring the computational account of the structure of the reservoir, and as we will see later, this point is important for PRC.

The second misconception is that the reservoir weights should remain unchanged, and experimenters cannot tune them in any sense.
This is untrue.
This mistake is thought to be raised from the expression of the RC learning scheme that the training is performed in the readout part.
This expression of the learning scheme is true, but this does not always mean experimenters cannot tune the weights of the reservoir.
An obvious counterexample is that when setting the ESN, it is common to tune the spectral radius of the reservoir weights \cite{Jaeger1,Jaeger_tutorial,Jaeger2,Schrauwen1,ESN_design}.
This is nothing but the tuning, or preconditioning, of the internal weights before training the readout to some specific task.
Another example can be found in cases that implement pretraining in the reservoir part before training the entire system for some specific target task.
The use of recurrent infomax \cite{Tanaka_infomax}, which maximizes the mutual information between the past and future within the internal dynamics, or the implementation of the plasticity rule, such as Hebbian learning \cite{LSM_Hebbian} or spike-timing-dependent plasticity (STDP) \cite{LSM_STDP1,LSM_STDP2}, into the input-driven RNN have been reported in pretraining the reservoir.
From this viewpoint, the recently introduced RNN called ALBERT \cite{ALBERT} for language processing can be included as a pretrained reservoir whose internal networks are pretrained based on predicting the ordering of two consecutive segments of text in the language data set; here, the readout part is trained for specific language-processing tasks.

The third misunderstanding is that if the reservoir is exhibiting chaos, which is a frequently observed behavior of nonlinear dynamical systems, then this means it cannot be used successfully.
Chaos can be characterized by sensitivity to initial conditions, where a slight initial difference in the state expands exponentially, and in this sense, the current state of the system is certainly affected by the initial condition.
Accordingly, although the chaotic dynamics show a rich diversity of patterns for function emulation, it seems that chaos does not show ESP and is not suitable for RC.
However, this is not the case.
Even if the dynamical system exhibits chaos, when it is driven by the input sequence (or noise), chaos is sometimes suppressed, and generalized synchronization occurs between the input sequence and the response of the dynamics \cite{Toral_NIO}, which is an outcome of ESP.
In particular, chaos in a large ESN equipped with a sigmoidal function \cite{Sompolinsky_Chaos} can be suppressed with noise \cite{Supress_Chaos}.
There exists a learning scheme that exploits this property of chaos suppression effectively, and it is found in the study of a first-order-reduced and controlled-error (FORCE) learning approach \cite{FORCE}.
In the study of FORCE learning, it was found that a chaotic reservoir is capable of implementing coherent patterns by adjusting the readout weights with the output fed back to the reservoir, or interestingly, the learning performance was even better than a non-chaotic reservoir in this condition.
Furthermore, because there is no fundamental difference between the output node fed back to the reservoir and the reservoir nodes interacting with each other, both through linear connection weights (although there is a slight difference concerning whether the output is injected into the nonlinear activation function before fed back to the reservoir), the FORCE learning scheme has been applied not only to the readout weights, but also to the internal weights of the reservoir \cite{FORCE,FORCE_transferring,FORCE_general}.
Chaos is more apparently exploited in the learning scheme, which is called innate training \cite{Innate}.
Chaos has rich dynamics but does not guarantee reproducible input--output relations.
Then, why not keep the richness of the dynamics and make it reproducible?
In the innate training approach, preparing the chaotic reservoir at first and collecting its own chaotic dynamics as training data, the internal connection weights are trained using FORCE learning to output their own chaotic dynamics reproducibly.
This approach can be also viewed as pretraining of the reservoir and has been applied for several machine learning and robot control tasks (see, e.g., Ref. \cite{Innate_ML,Inoue_Inputreservoir,Inoue_CI}).
Recently, many neuromorphic devices have been shown to exhibit chaos (e.g., Ref. \cite{Morie_chaos,Spin_Chaos1,Spin_Chaos2}), and it is expected that these chaotic dynamics can be harnessed and exploited as a computational resource based on an RC framework.   

Because the tuning of the internal weights were introduced in the above approaches, it may be helpful to clarify the difference between the conventional training scheme, such as BPTT, and the above introduced approaches.
The main difference comes from the design of the cost function.
In BPTT, there usually exists a global target function, and the gradient is obtained based on it; in addition, the error is backpropagated to each internal node to be used to update the concerning weights.
In the above approaches, however, the internal weights are not usually tuned for the global target function but can be tuned for any global or local target function that the experimenter designs.
In this sense, the above approaches contain more freedom in the setting of cost functions, or it may be more appropriate to say that these approaches even include the conventional setting of cost function.
This RC property, which can be composed of multiple cost functions, is also an important aspect to be kept in mind when trying to step toward the PRC.

In this paper, we discuss what becomes interesting when we proceed from conventional RC driven inside a PC (this is also physical dynamics, though) to PRC that exploits physical dynamics as a reservoir.
The story begins from the genesis of LSM, which is one of the original RC model systems.

\section{Liquid State Machine}
\subsection{``Wetware'' and its implication}
When Wolfgang Maass, Thomas Natschl\"{a}ger, and Henry Markram proposed the seminal model of the LSM, at around the same time, Wolfgang Maass presented some interesting insights in his paper entitled ``Wetware'' about the modality of information processing in the brain \cite{Maass_Wetware}.
This paper starts as follows: \\
\newline
{\it 
``If you pour water over your PC, the PC will stop working. 
This is because very late in the history of computing – which started about 500 million years ago – the PC and other devices for information processing were developed that require a dry environment. 
But these new devices, consisting of hardware and software, have a disadvantage: they do not work as well as the older and more common computational devices that are called nervous systems, or brains, and which consist of wetware. 
These superior computational devices were made to function in a somewhat salty aqueous solution, apparently because many of the first creatures with a nervous system were coming from the sea. 
We still carry an echo of this history of computing in our heads: the neurons in our brain are embedded into an artificial sea-environment, the salty aqueous extracellular fluid which surrounds the neurons in our brain. ...''
}
\newline
\\
Maass's paper \cite{Maass_Wetware} subsequently discusses how to capture the information processing function of the human brain.
The idea expressed in the above introductory paragraph already penetrates the fundamental aspect of PRC.

The important point that we should confirm here is that once computation, which is an abstract input—output operation in principle, was implemented in the real-world through a physical entity or substrate, then the physical property of the substrate and the influence of its execution environment came to affect the implemented computation and inevitably added a novel property/functionality to the system. 
The above example clearly suggests that even if the same computation is implemented, according to the choice of physics for the substrate (in the above case, the conventional PC and brain), the robustness against water is different.

The conventional PC consists of hardware and software; the hardware is the ``physical'' part of the PC, and the software is a set of commands used to run it. 
These two components function complementarily. 
That is, the hardware is specialized and designed to execute the command sent from the software. 
In contrast, the nervous system can function in a somewhat salty aqueous solution, but this physical condition is not fully designed for information processing. 
Rather, the nervous system exploits its given environmental constraints and physical conditions—which are shaped by its original context (i.e., many of the first creatures with a nervous system came from the sea)—to enable information processing.

When we look at the background of the seminal model of the LSM, it is evident that Wolfgang Maass and his colleagues were not adopting a conventional view of the brain as a network consisting of interacting elements (i.e., neurons) as many researchers do; instead, they characterized its behavior based on the surrounding liquid physical substrate. 
Furthermore, the idea is not merely a metaphor; the researchers even proposed a concrete sketch of their proposed model. 
This system is called the ``liquid computer.'' \cite{Fluid_Maass}

\subsection{Liquid computer and the liquid brain}
Thomas Natschl\"{a}ger, Wolfgang Maass, and Henry Markram suggested that the brain is constantly exposed to a massive flow of sensory information, including both audio and visual inputs, and that it does not exist in the stable state frequently expressed as an attractor but rather in a transient state (except when it is in the ``dead'' state) \cite{Fluid_Maass}. 
According to this view, they proposed a scheme to exploit the surface of a liquid (such as a cup of coffee) for computation.

We focus on a transformation over a time series. 
We consider the issue by mapping from input sequences $\mbox{\boldmath $u$}(\cdot)$, which are a function of time, to output sequences $\mbox{\boldmath $v$}(\cdot)$, which are also a function of time; this transformation is usually called a filter (or operator).

\begin{figure*}
\includegraphics[width=150mm]{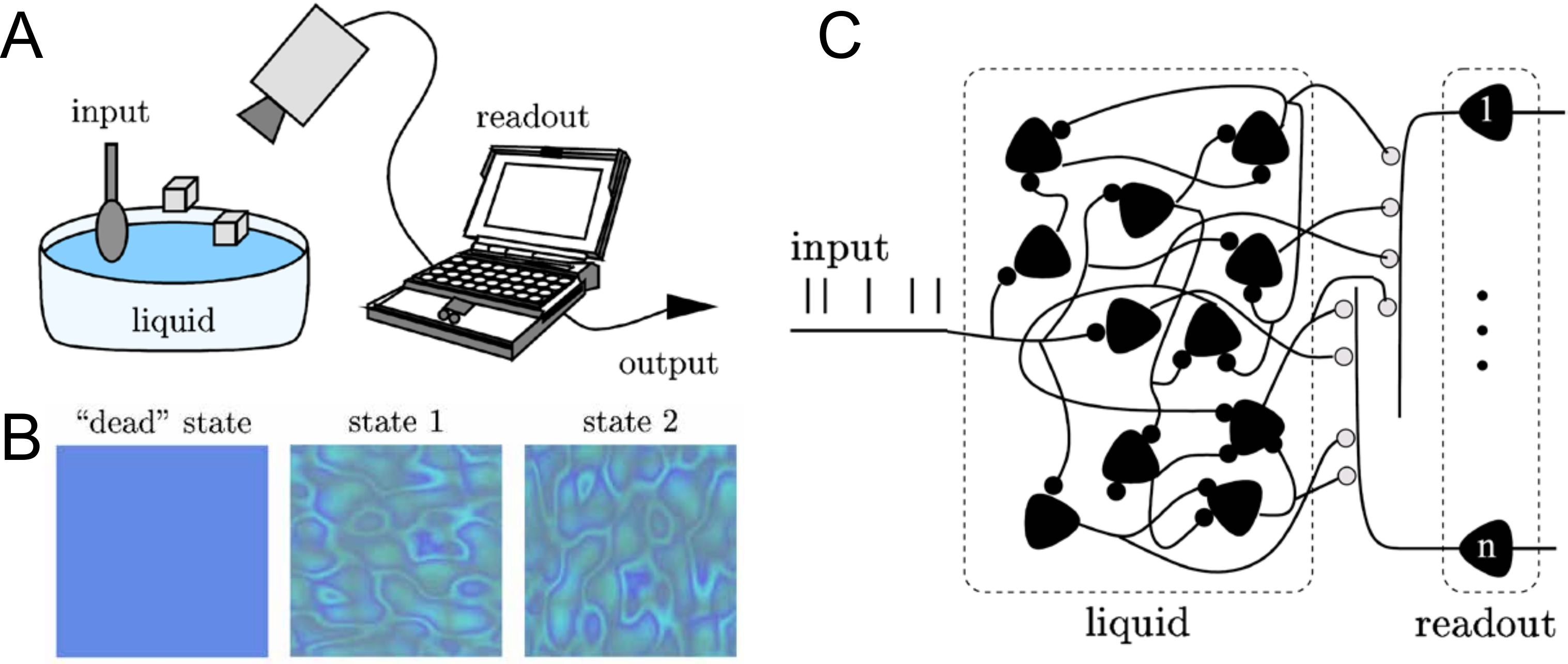}
\caption{{\bf A Natschlager-Maass-Markram-type liquid computer and its analogy to neural information processing.}
A. Schematics of a ``Liquid computer.''
Figure reprinted with permission from Ref. \cite{Fluid_Maass} by the author.
B. The system takes a video image of a liquid surface as a state of the system. 
The liquid surface shows different spatiotemporal patterns according to how it is perturbed (e.g., manual perturbations using a spoon or dropping a cube of sugar, and their temporal orderings make the patterns of the liquid surface different, such as ``state 1'' and ``state 2'').
Figure reprinted with permission from Ref. \cite{Fluid_Maass} by the author.
C. Understanding neural circuits as an LSM.
Figure reprinted with permission from Ref. \cite{Fluid_Maass} by the author.
}
\label{fig4}
\end{figure*}

See Fig. \ref{fig4}A. 
The schematics show the conceptual design of a ``liquid computer.'' \cite{Fluid_Maass}
To illustrate this concept, one could imagine a situation where he or she prepares a cup of coffee and perturbs the coffee surface by using a spoon or dropping a cube of sugar in, thereby injecting an ``input.''
Consider that we have a video camera that can monitor the coffee's surface in real time and define this camera image at time $t$ as the liquid state $\mbox{\boldmath $x$}(t)$ (Fig. \ref{fig4}B).
The liquid (in this case, coffee) transforms the input time series $\mbox{\boldmath $u$}(\cdot)$ into a liquid state $\mbox{\boldmath $x$}(t)$, expressed as $\mbox{\boldmath $x$}(t)=(L\mbox{\boldmath $u$})(t)$, where $L$ is called a liquid filter.
The image is sent to the PC, and by using the state of the surface, the PC processes the state and outputs the result.
The interesting point of this system is that one can design various filters without using the memory storage inside the PC; in other words, this process can be carried out with the memory-less readout $f$, expressed as $\mbox{\boldmath $v$}(t)=f(\mbox{\boldmath $x$}(t))$. 

Let us consider an example of information processing using this system. 
Assume that we want the system to output the number of cubes of sugar injected over the last two seconds. 
Because the readout part in the PC is memory-less, to perform this task, the current liquid state should be able to express the number of cubes of sugar droped inside over the last two seconds in a distinguishable form. 
Let us call this ability to distinguish the previous input state as a difference in the current liquid state the ``separation property'' of the liquid. 
Then, to perform the task, it is necessary to map the separated states into the required output (e.g., the liquid states perturbed in the order of ``$spoon \rightarrow cube \rightarrow cube$'' and ``$cube \rightarrow spoon \rightarrow cube$'' should be mapped to output ``$2$''). 
This property of the readout function is called the ``approximation property.'' 
Interestingly, it has been shown that any time invariant filter with fading memory can be approximated in an arbitrary precision composing these two properties (with a filter bank containing point-wise separation property and readout function having universal approximation property) (see, Ref. \cite{Maass1,Fading_memory,Maass_theory} for detailed discussions). 
In Ref. \cite{Fluid_Maass}, it is stated that the formalization of this liquid computer is an LSM and is proposed to understand the information processing of a neural circuit (Fig. \ref{fig4}C).

\begin{figure}
\includegraphics[width=80mm]{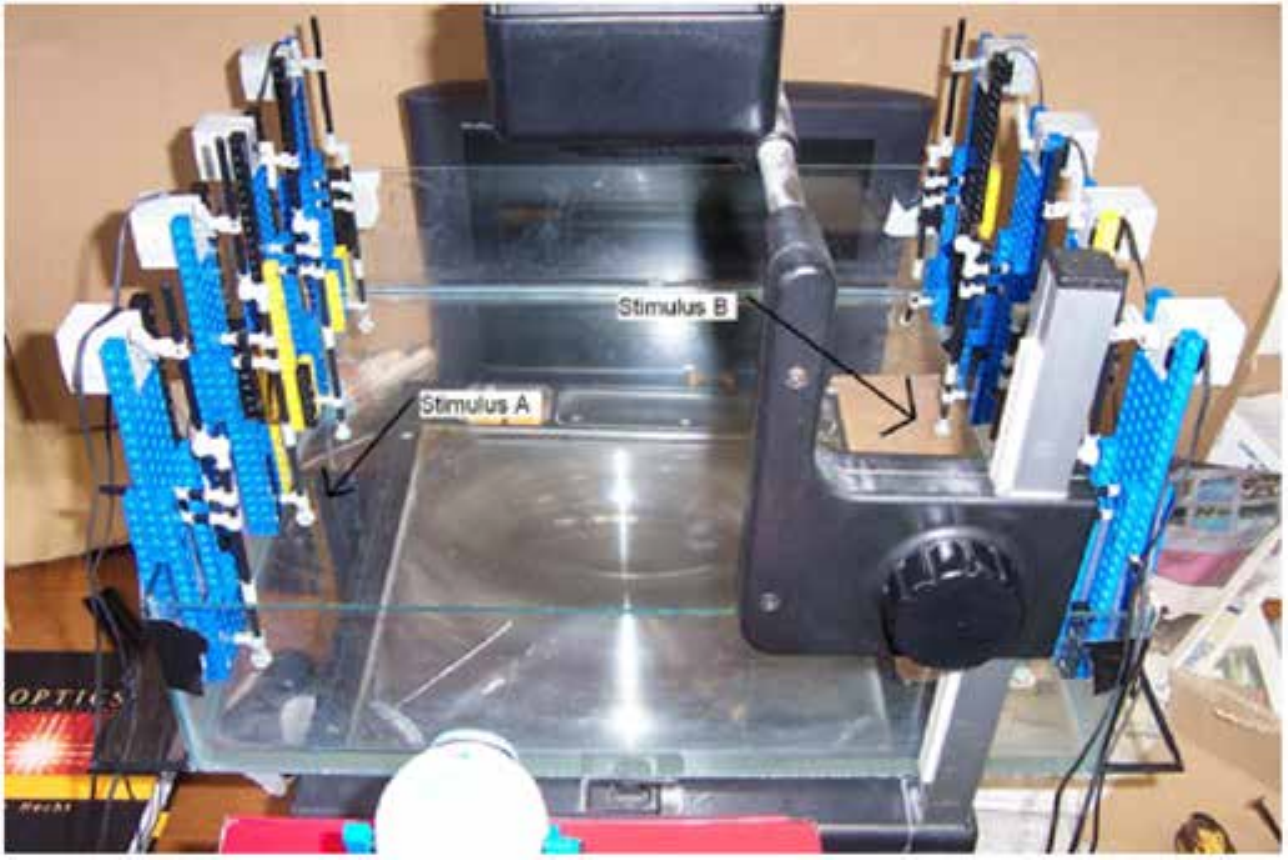}
\caption{{\bf A Fernando-Sojakka-type liquid brain.}
Figure reprinted with permission from Ref. \cite{Fluid_Fernando}, Copyright (2003) by Springer Nature.}
\label{fig5}
\end{figure}

As soon as the concept of the liquid computer and its formalization under an LSM were proposed, two computer scientists, Chrisantha Fernando and Sampsa Sojakka from the University of Sussex, integrated the idea into a physical system; they called this model the ``liquid brain'' (Fig. \ref{fig5}). 
In their paper \cite{Fluid_Fernando}, they described it as follows: \\
\newline
{\it 
``
... Here we have taken the metaphor seriously and demonstrated that real water can be used as an LSM for solving the XOR problem and ...
''
}
\newline
\\
Using this system, they showed that water in a bucket is capable of implementing an XOR task and a speech recognition task \cite{Fluid_Fernando}. 
We can confirm that water in a bucket is not made or designed for computation but can be exploited for it.
(Note that, recently in complex systems study, coginitive networks that lack stable connections and static elements are also called liquid brains \cite{liquid_solid_brain}.
These networks include such as ant and termite colonies, immune systems, and slime moulds.)

\section{Soft Robotics}
A computer is, in simple terms, a machine that is made to compute. 
Accordingly, the hardware structure of a computer is specialized to implement computation in general. 
Here, the concrete form of computation is determined beforehand in a top-down manner, and to realize it, the component arrangement is designed and decided in detail. 
On this point, the liquid computer and liquid brain are composed in somewhat opposite directions compared with the conventional computer. 
They both started from the physical property of liquid, and by considering how to exploit this property for computation, they came to invent a novel scheme to implement it, which is a bottom-up approach.

\subsection{Embodiment and morphological computation}
In robotics, a concept that accounts for these unexpected and intrinsic properties associated with the physical body when implementing computations, abstract operations, or behavior control has been around for a long time. 
This concept is called ``embodiment.'' \cite{Rolf_Understanding,Rolf_How,Rolf_Science}
For example, a seminal platform called a ``passive dynamic walker'' can walk naturally like a human without having an external controller \cite{PDW}. 
Just by using a well-designed body (a compass-like shape) and a well-designed environment (a slope), the natural walking behavior can be realized, where the behavior control is partially outsourced to the physical body. 
In bio-inspired robotics, this property of embodiment is studied in various platforms, including not only bipedal walkers, but also quadruped robots (e.g., Ref. \cite{Qian_ROBIO,Qian_IROS2012,Nico2013,Qian_AdvancedRobotics}). 
Similar properties can be found in animals. 
There is a famous experiment that used the dead body of fish (a trout, specifically) where the body was able to generate a vivid and natural swimming motion by exploiting the vortex in a water tank \cite{DeadFish}. 
In this experiment, because the fish was dead, we can guarantee that the central nervous system of the fish was not functioning at all, so we can also confirm that the specific morphology and material property of the body and its interaction with the vortex in the surrounding water environment were capable of realizing the natural swimming motion of a fish \cite{DeadFish}. 
In the field of self-assembling systems, there are many studies that investigate how the shape of each element induces or affects the global behavior of the system (e.g., Ref. \cite{SelfAssembling1,SelfAssembling2,SelfAssembling3,SelfAssembling4}). 
Here, the research field that aims at investigating and pursuing the nature of how the shape or morphology of the system affects the behavior of the entire system is called {\it morphological computation} \cite{Morphological_Computation}.

\begin{figure*}
\includegraphics[width=170mm]{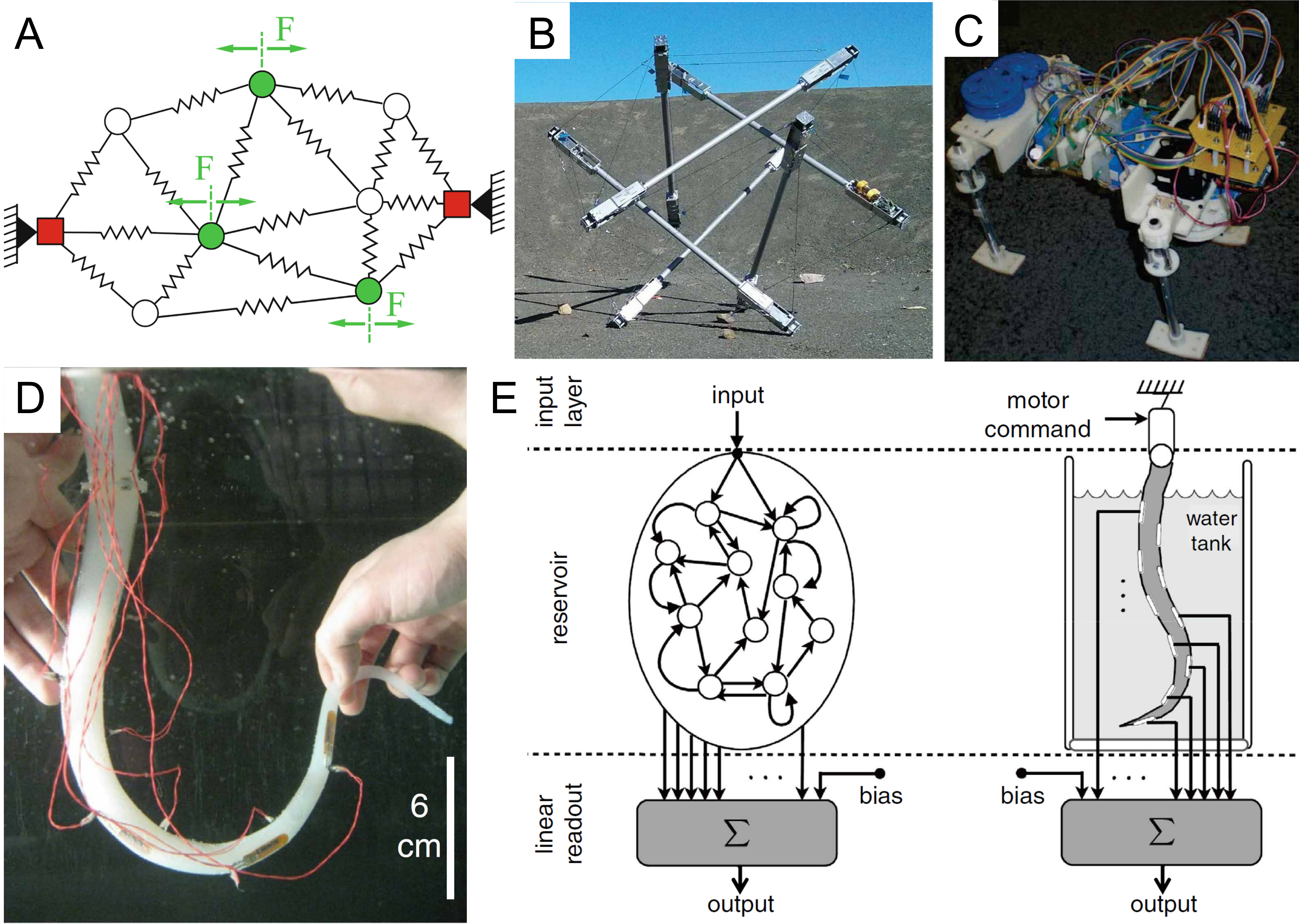}
\caption{{\bf Physical reservoir computing using compliant and soft bodies.}
A. A generic mass-spring network used as a reservoir in Ref. \cite{Mech_Helmut1}.
Figure reprinted from Ref. \cite{Mech_Helmut1} under the Creative Commons CC-BY-NC license.
B. A tensegrity robot called SUPERball proposed in Ref. \cite{Nasa_superball}.
Figure reprinted with permission from Ref. \cite{Nasa_superball}, Copyright (2015) by IEEE.
C. A quadruped robot called Kitty proposed in Ref. \cite{Mech_Qian}, which exploits soft spine dynamics as a reservoir.
Figure reprinted with permission from Ref. \cite{Mech_Qian}, Copyright (2013) by IEEE.
D. A picture of a physical soft robotic arm inspired by the octopus used in the experiment in Ref. \cite{Mech_Nakajima4}. 
It is made of silicone and embeds ten bending sensors, monitoring the soft body dynamics every 0.03 [s].
Figure reprinted from Ref. \cite{Mech_Nakajima4} under the Creative Commons license.
E. Schematics explaining how to exploit the soft robotic arm as a reservoir.
Figure reprinted from Ref. \cite{Mech_Nakajima4} under the Creative Commons license.
}
\label{fig6}
\end{figure*}

Are there any quantitative ways to characterize the intrinsic information processing capability of the physical body? 
Helmut Hauser et al. tried to propose a framework to theoretically investigate the morphological computation of compliant bodies \cite{Mech_Helmut1}. 
In their study, they considered a mass-damper system, which is often used to model the body of robots, and explained that by using a linear mass-damper system, it is possible to compose a filter bank, which we discussed earlier (Fig. \ref{fig6}A). 
This implies that if you design the readout function nicely, it is possible to approximate time-invariant filters with a fading memory property using a linear mass-damper system, which is consistent with the arguments corresponding to the LSM model. 
Furthermore, the authors numerically demonstrated that by using a complex nonlinear mass-damper system, even the nonlinearity required in the readout function can be outsourced to the mass-damper system, and the system would be capable of emulating nonlinear filters with fading memory only by composing the linear readouts. 
That is, this approach suggests that the physical body of robots can be, in some conditions, used to emulate nonlinear filters with a fading memory, which implies that the physical body can be used as a successful reservoir.
Subsequently, Helmut Hauser et al. have investigated the role of feedback on the mass-damper system implementing nonlinear limit cycles based on this framework \cite{Mech_Helmut2}.
Tensegrity structures serve as an appropriate testbed to implement this framework, where it enables the structures to embed closed-loop control and realize locomotion by exploiting its intrinsic body dynamics as a computational resource, here being a controller \cite{Mech_Ken1,Mech_Ken2,Mech_Wyffels}  (Fig. \ref{fig6}B).
(Note that, although we do not go into details in this paper, information theoretic approachs to characterize morphological computation have also been investigated \cite{MC_Info0,MC_Info1}.)

\subsection{Physical reservoir computing using a soft robotic arm}
Soft robotics is a recently developed field that actively investigates the account of soft and compliant bodies to functionality and behavioral control \cite{SoftRobots1,SoftRobots2,SoftRobots3}.
Compared with conventional rigid-bodied robots, soft robots introduce a number of novel challenges into the field regarding the material properties and deformable morphology of the body as well as the complexity and diversity of body dynamics.
Soft robots hold many advantages, which are linked to the mechanical softness of the body \cite{SoftRobots1,SoftRobots2,SoftRobots3,Li_Octopus}.
For example, they are considered to be useful in the situation of human–robot interaction, rescue, and biomedical applications because they do not damage people in the same way that rigid robots do; in other words, they are generally considered a safer option.
These robots, however, include challenges in terms of control \cite{Li_Octopus}.
Soft robots are often classified into the category of an underactuated system, where the number of the actuation points are less than the degrees of freedom.
Furthermore, they usually generate diverse and complex body dynamics when actuated, which are high-dimensional, nonlinear, and contain short-term memory \cite{Nakajima_info1,Nakajima_info2,Nakajima_info3}.
These properties make soft robots difficult to control using the conventional control scheme.

On the other hand, these seemingly undesirable properties of soft robot control can be viewed as a positive from PRC perspectives.
That is, we can exploit the diverse, rich dynamics of a soft body as a computational resource—more specifically, as a reservoir (Fig. \ref{fig6}C, D, and E).
In previous studies, we have shown that a silicone-based soft robotic arm inspired by an octopus can be used as a successful reservoir by taking the actuation sequence as the input and sensory reading as the reservoir state; indeed, this method exhibits high information-processing capability in some conditions \cite{Mech_Nakajima3,Mech_Nakajima4,Mech_Nakajima6,Mech_Nakajima7}  (Fig. \ref{fig6}D).
Interestingly, octopus arms have characteristic muscle organizations termed muscular-hydrostats \cite{Soft_Octo}.
In these structures, the volume of the organ remains constant during their motion, enabling diverse and complex behaviors.
We showed that using biologically plausible parameter settings, the dynamic model of the muscular-hydrostat system has the computational capacity to achieve a complex nonlinear computation \cite{Mech_Nakajima1,Mech_Nakajima2,Mech_Nakajima5}.
Furthermore, by incorporating the feedback-loop from the output to the next input (i.e., the next actuation pattern), we have demonstrated that the robot's behavioral control for the next time step can be implemented by using its current state of its body as a computational resource, suggesting that the ``controller'' and ``to be controlled'' is the same in this scheme \cite{Mech_Nakajima3}.
This concept has been also applied to the study of a quadruped robot, where the robot exploits its spine dynamics as a physical reservoir to control its actuation patterns and locomotions \cite{Mech_Qian} (Fig. \ref{fig6}C). 
In short, the drawbacks of soft robot control became assets for control from a PRC viewpoint.

\section{Exploiting physical dynamics for computational purposes}
We began by reviewing the concept of wetware by Wolfgang Maass, and from there, we illustrated the development of physical platforms, such as the liquid computer, liquid brain, mass-damper systems, and silicone-based soft robotic arms inspired by octopuses. 
In this section, we would like to review three significant phases that we can find in this evolution.

\begin{figure*}
\includegraphics[width=160mm]{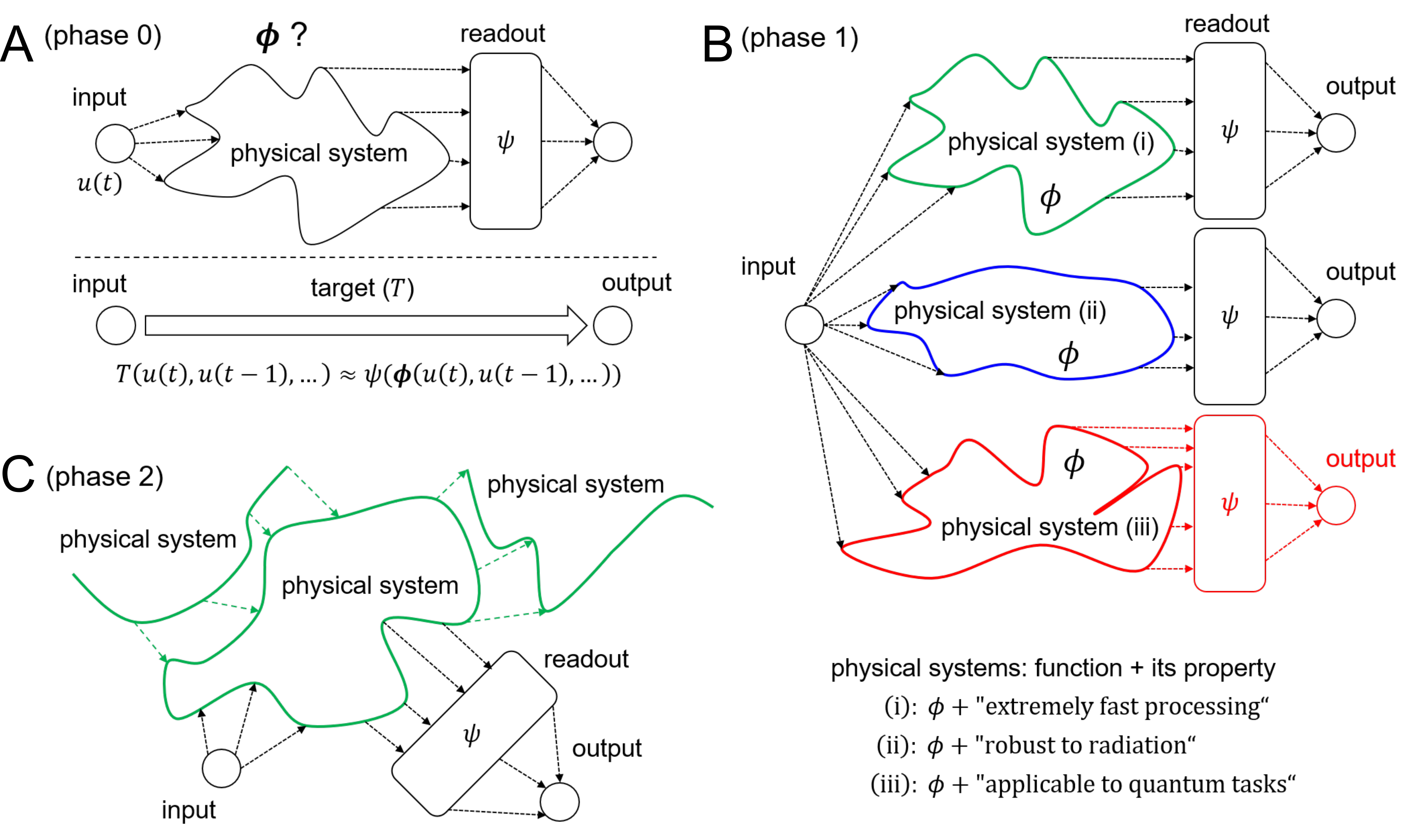}
\caption{{\bf Three phases in PRC.}
A. Phase 0.
PRC can be used as a method to infer the information processing capability of natural physical dynamics.
B. Phase 1.
Physical properties, which are potentially different according to the type of physics, are added to the reservoir in PRC.
C. Phase 2.
PRC enables to exploit physical dynamics as a computational resource that is already functioning for different purposes.
See the text for details.}
\label{fig7}
\end{figure*}

\subsection*{Phase 0: Inferring the computational power of physical systems.}
PRC provides a method to exploit natural physical dynamics as a computational device.
It implies that this method is also useful for investigating which physical systems are suitable to implement which types of computation and for analyzing the information-processing capability of the physical dynamics.
In particular, if we use linear and static readouts to generate outputs for specific tasks requiring a certain amount of nonlinearity and memory, because we are not adding any nonlinear terms and memory externally, by evaluating the task performance, we can infer back which amount of nonlinearity and memory has been positively contributed or exploited from the physical reservoir to perform the task (Fig. \ref{fig7}A).
That is, in this way, if we use a previously introduced symbol, we can pursue the nature of the function $\mbox{\boldmath $\phi$}$ in the physical systems.
Systematic investigations are needed to reveal the response characteristics of the physical system against the type, intensity, and timescale of the input, and these properties are intrinsic to each physical system.
Accordingly, we can expect the diversity of the type of information processing according to the type of physics, where each physical system has a preference in terms of the type of functions it can express.

In neuroscience, there are several studies that have inferred the computational capability of the neural circuits \cite{Bio_Maass} or the cultured neural systems \cite{Bio_Dranias,Bio_Kubota}.
Obviously, their motivation is not to make a high-performance computer but rather to reveal the functional characteristics of the natural systems from information-processing perspectives.
This approach can also be applied to infer the functionality of the body of living systems quantitatively.
As we discussed in the concept of embodiment, a functionality that is thought to be handled by the brain is often partially outsourced to the physical body.
Unlike the randomly coupled ESN, the biological body has a specific structure or morphology that is intrinsic to respective living organisms.
This specific morphology is evolved through the respective ecological niche of living things, which is a driving force of the diversity of morphology.
It is expected that the PRC framework has the potential to reveal the property of the body's morphology from information-processing perspectives.
(Related to this issue, there exists a research project that aims to characterize RC from evolutionary perspectives \cite{RC_evolution}.)
The above directions of research can be summarized and stated as the study of $\mbox{\boldmath $\phi$}$ within the physical system.
This penetration is the basics and is fundamentally important in the PRC framework and can thus be taken as a ground basis, which we call phase 0.

We should note, however, that once the function $\mbox{\boldmath $\phi$}$ of the physical system is revealed, then because it is a mathematical description in principle, there is no meaning to use the actual physical system as an information-processing device anymore, but we can implement the same functionality of the physical system using a conventional PC.
If we only stick to this perspective, then PRC does not differ so much from the original RC anymore.
Shortly, the diversity we can find here is in fact the diversity of function $\mbox{\boldmath $\phi$}$.
Now, much like as we overviewed from the examples starting from wetwares, PRC has the potential to go beyond this perspective.
That is, the PRC framework can deal with a property that is not described in $\mbox{\boldmath $\phi$}$. 
This point is elaborated subsequently in phase 1 and phase 2.

\subsection*{Phase 1: Physical properties of a computer.}
A computer is a machine that is designed for computation.
As long as it is made of a physical entity, it inevitably and, sometimes unexpectedly, adds physical and material properties to the system that are not always directly connected to the computational purpose (these properties can present as both advantages and disadvantages for the user).
We have clearly confirmed this point using the example of wetware, and this constraint is also true for PRC.
Even if you implement the same computation, depending on the type of physics you exploit, you may gain additional or unexpected properties beyond the computation itself (Fig. \ref{fig7}B).
For example, if you use a laser as a computational resource, you can implement an extremely fast computation, or if you use water as a substrate, the system will be tolerant of water (Fig. \ref{fig7}B).
Many currently discussed assets of physical reservoirs can be understood from this perspective.
In particular, spintronics devices have been gaining attention as an appropriate substrate for PRC because of their compactness, high-speed processing, and energy efficiency while being able to function at normal temperatures \cite{Spin_Julie1,Spin_Furuta_PRAppl,Spin_Tsunegi1,Spin_Nakane1,Spin_Tsunegi2,Spin_Julie2,Spin_Nomura1,Spin_Kanao,Spin_Julie3,Spin_Liu,Spin_Nomura2,Spin_Julie4}.
These assets are somewhat common in the computer science field, but spintronics devices also contain an interesting additional property: they show high durability in radioactive environments \cite{Spin_radioactive} (Fig. \ref{fig7}B).
This property opens up the potential for spintronics reservoirs to be used as a computational substrate in extreme environments where conventional electronic devices break down or do not function at all.
Another example can be found in quantum reservoir computing.
Since the first conception to exploit quantum dynamics as a reservoir in Ref. \cite{QR_Nakajima1}, there have been many variants and extensions proposed in the literature \cite{QR_Nakajima2,QR_Ghosh1,QR_Ghosh2,QR_Ghosh3,QR_Yamamoto}.
In quantum reservoir computing, by using the property of quantum computational supremacy, a huge amount of computational nodes can be equipped, which then provide a direct influence to the information-processing capability of the system \cite{QR_Nakajima1,QR_Nakajima2}.
Another important property is that because quantum reservoir computing exploits quantum dynamics, it is capable of implementing a quantum task (a task defined in the quantum scale) (Fig. \ref{fig7}B).
In Ref. \cite{QR_Ghosh2}, the preparation of desired quantum states, such as single-photon states, Schr\"{o}dinger's cat states, and two-mode entangled states, is introduced as an effective application domain. 

To induce these assets, which originate from the physical properties of a reservoir, current technologies still require conventional electronics and external devices, such as for the readout part, to maintain the temperature during the reservoir executions and to make the physical reservoir work in the real environment.
This is a weakness of currently available technologies; these points should be improved, and a novel scheme should be proposed in the future.

\subsection*{Phase 2: Exploiting a physical substrate that is not made for computation for computation.}
\begin{figure}
\includegraphics[width=85mm]{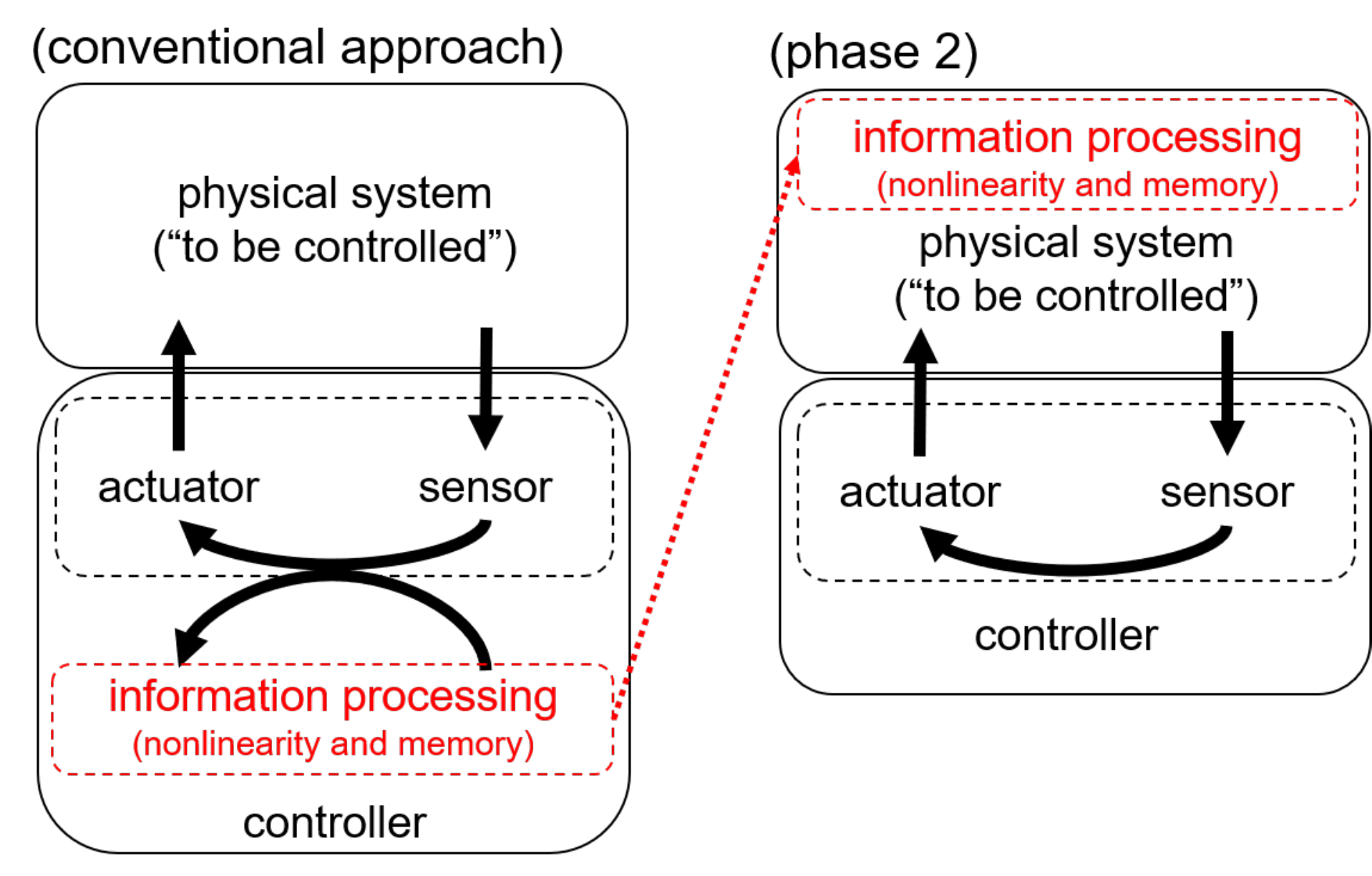}
\caption{{\bf Schematics of the closed-loop control in phase 2 of PRC.}
In conventional control, information processing is prepared outside a system that is to be controlled or acted on (left diagram).
In phase 2, information processing is accompanied by the behavior of the system (right diagram). 
The system behaves in a certain manner and performs information processing simultaneously. 
Note that the required information processing to generate behavioral control can be bypassed from the digital processor and embedded in the system itself.}
\label{fig8}
\end{figure}

If we think of the body of a robot, it is, of course, not made for computation.
The body is an essential constituent of a robot and is inevitably associated when generating behaviors.
That is, the robot's intended functionality is to realize behaviors in the real world.
As we have seen in the example of soft robots, if the body itself exerts certain dynamic conditions, then according to the PRC framework, the body can also be used as a computational resource (Fig. \ref{fig7}C).
This implies that the two functionalities—``behavioral generation'' and ``information processing''—are associated with the same physical body.
Then, when the robot generates behavior, we can simultaneously use its dynamics for information processing.
Considering this property, as we confirmed in the above examples of soft robotic arms, together with an incorporation of the feedback-loop, the resulting body dynamics of the target behavior can be exploited to calculate the target motor command that controls its own behavior.
It shows that this approach is more efficient than any other controller attached externally to realize target robotic behavior (Fig. \ref{fig8}).

Phase 2 may be classified as a derivative of Phase 1, but the major turn from Phase 1 to Phase 2 is that in the latter, the physical substrate is not prepared for computational purposes whatsoever in the first place.
The most interesting point of PRC among other computational frameworks can be found here.
PRC can easily generate the transition from Phase 1 to Phase 2.
This is because the RC framework allows one to exploit the natural dynamics of physical systems for information processing.
Accordingly, in PRC, we do not need to precisely design the physical substrate specific to target computation in many cases; rather, the implemented information processing depends on the input-driven dynamics of the physical substrate, which results in a diversity of information processing. 

Then, which kind of physical reservoirs is classified in Phase 2 other than a soft robotic arm inspired by an octopus?
This question is a fundamental theme that should be further explored in the field of PRC. 
One direction would be to exploit real living things, such as animals (e.g., rats, fish, etc.) or the human brain, as a physical reservoir. 
In principle, living things are free from the intended purposes introduced by users. 
Needless to say, they are not made for computational purposes. 
Recently, there have been several studies suggesting that the brain wave of animals and humans exhibit consistent responses against external inputs, and it is expected that the PRC approach can be directly applied to brain waves (e.g., Ref. \cite{Brainwave_Consistency}).
This direction of research has long been studied in the field of brain-machine-interface.
Together with the recent advancement of sensing technology that allows us to monitor massive amounts of data from living things (e.g., Ref. \cite{Massive_Signal_Brain}), the PRC approach presents a high potential for further study of the issue, and it can be actively applied not only to our daily devices, such as smart phones, but also to wearables and biomedical devices.

\section{Acknowledgements}
K. N. would like to acknowledge Taichi Haruna, Katsushi Kagaya, Megumi Akai-Kasaya, Atsushi Uchida, Kazutaka Kanno, Sumito Tsunegi, Quoc Hoan Tran, and Yongping Pan for their fruitful discussions and thoughtful suggestions.
This work was based on results obtained from a project commissioned by the New Energy and Industrial Technology Development Organization (NEDO). 
K. N. was supported by JSPS KAKENHI Grant Numbers JP18H05472 and by MEXT Quantum Leap Flagship Program (MEXT Q-LEAP) Grant Number JPMXS0118067394.

\end{document}